\documentclass[%
reprint,
nofootinbib,
amsmath,amssymb,
aps
]{revtex4-2}

\usepackage{subcaption}

\usepackage{graphicx}
\usepackage{dcolumn}
\usepackage{bm}
\usepackage{amsmath}
\usepackage{mathtools}

\usepackage{graphicx}
\usepackage{dcolumn}
\usepackage{bm}
\usepackage{amsmath}
\usepackage{mathtools}

\begin{document}

\title{ Probing the mesoscopics of competing interactions with the thermodynamic curvature: the case of a two-parameter ANNNI chain.
}

\author{Soumen Khatua}
\email{soumenk.ph21.ph@nitp.ac.in }
\affiliation{National Institute of Technology Patna}

\author{Anurag Sahay}
\email{anuragsahay@nitp.ac.in}
\affiliation{National Institute of Technology Patna}

\begin{abstract}

This work examines the full scope of long-standing conjectures identifying the invariant thermodynamic curvature $R$ as the correlation volume $\xi^d$ and also as a measure of underlying statistical interactions. To this end, we set up a two-parameter ANNNI (Axial Next Nearest Neighbour Ising) chain featuring two next nearest neighbour (nnn) and a nearest neighbour (nn) interaction. Competition between interactions and resulting frustration engender a rich phase behaviour including a cross-over between two ferrimagnetic sub-phases. We show that $R$ attests to all its conjectured attributes with valuable insights into the character of mesoscopic fluctuating substructures. In a remarkable demonstration of its relevance  at a far-from-critical point, $R$ is shown to resolve a hitherto unnoticed tricky issue involving $\xi$. A physically transparent expression for the zero field $R$ helps bring into focus the pivotal role played by some third order fluctuation moments. 

\end{abstract}

\pacs{Valid PACS appear here}
\maketitle

{\textit{Introduction}}: Macroscopic thermodynamics emerges as a weighted contribution of all allowed microscopic configurations. As such it coalesces information about all possible $n$-point correlations, only some combinations of which may be usefully recovered by classical thermodynamic fluctuation theory (CFT) and its Gaussian approximation. For example, the magnetic susceptibility, representing a sum over all two-point correlations, is useful in characterizing phase transitions.
Higher order cumulants, though relevant to simulations and advanced statmech calculations, find limited use in a macroscopic set-up.

Thermodynamic geometry, a covariant extension of CFT, can potentially recover some microscopic information in a consistent manner, \cite{rupporiginal,rupprev}. An element of geometry, the scalar curvature $R$ is expressed as an invariant combination of third order and second order moments. Significantly, it is found to be  equal to the correlation volume $\xi^d$ near criticality, modulo an order unity constant. Hyperscaling arguments equate it to the inverse of singular free energy, $R=\kappa\psi_s^{-1}$ leading to a thermodynamic description of the critical point. 

Remarkably, a heuristic understanding of $R$ as a typical size of a correlated domain (or Ruppeiner's $conjecture$ $R\sim\xi^d$, \cite{rupporiginal}) remains relevant even in non-critical regimes as evidenced by the geometric calculation of the sub-critical coexistence lines and the super-critical Widom line in simple fluids and magnetic systems, \cite{sahay1,may,sarkar1}. In addition, the sign of curvature is known to change from positive in solid-like or statistically repulsive fermionic states of aggregation  to negative in fluid-like or statistically attractive bosonic states, \cite{rupprep,behrouz2}.

In this work we take a definitive step forward in demonstrating the prowess of thermodynamic geometry at characterizing mesoscopic fluctuating structures in both near-critical as well as far-from critical regimes.  Importantly, we do so in an exactly solved model of competing interactions with rich meso-structures, where we directly verify the claims of geometry against exact microscopic correlation functions.

We introduce a generalized, two-parameter ANNNI chain where the nnn couplings between odd pairs and between even pairs of Ising spins can each independently range from anti-ferromagnetic to ferromagnetic while the nn coupling remains ferromagnetic. The ANNNI model, which follows as a special case, is  widely used to understand phase behaviour in systems with competing interactions and resulting frustration, \cite{elliot,steph,julia,selke,lieb}. 
Monte Carlo studies reveal a rich spatially modulated substructure of fluctuating and interacting domain walls and kinks, \cite{selke}. Our two-parameter chain exhibits an even more varied phase structure  which includes a ferrimagnetic phase in addition to ferromagnetic and antiphase states. Earlier works on spin lattices and frustrated systems include \cite{ruppmag,mrugala, dolan1,dolan2,dolan3,dolan4,brody,riek,alata,soum} and \cite{tala,belucci}.

\begin{figure*}[t!]
 \begin{subfigure}[b]{0.3\textwidth}
\centering
\includegraphics[width=3in,height=2in]{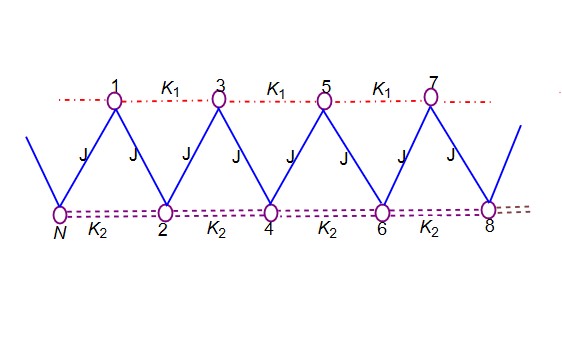}
 \caption{}
\end{subfigure}
\hspace{2.8cm}
 \begin{subfigure}[b]{0.3\textwidth}
\centering
\includegraphics[width=2.5in,height=2.5in]{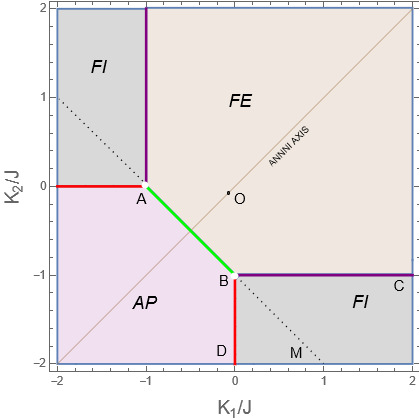}
 \caption{}
 \end{subfigure}
\caption{{\small{($a$) Two parameter ANNNI model as a series of frustrated triangles. ($b$) Ground state phase structure  in the $K_1K_2$ plane. }}}
\label{fig1}
\end{figure*}

\textit{The model and its phase structure}. The Hamiltonian of our generalized ANNNI chain is written as{\footnote{While pursuing this work we found a related and more general Hamiltonian in the interesting work \cite{plini}. Its context and parameter ranges are different from ours.}}
\begin{eqnarray}\mathcal{H}=&-&J\sum_{i}s_is_{i+1}- K_{1}\sum_{i=1,3,..}s_is_{i+2}\nonumber\\&-& K_2\sum_{i=2,4,..}s_is_{i+2}-H\sum_is_i
\label{Hamiltonian}
\end{eqnarray}

Figure \ref{fig1}(a) shows the two-parameter ANNNI chain (periodically identified for later use) as a series of frustrated triangles. The canonical ANNNI chain is recovered by setting $K_1=K_2$. To highlight the role of geometry we limit discussion to some salient aspects of zero field phase behaviour. A detailed survey follows in a future work.

Figure \ref{fig1}(b) shows ferromagnetic (\textit{FE}), ferrimagnetic (\textit{FI}) and antiphase (\textit{AP}) ground state configurations along with the phase boundaries in the $K_1K_2$ plane. The line marked BM is discussed later. Spin reflection symmetry lets us restrict to $H\to 0+$. $K_1\leftrightarrow K_2 $ symmetry across the ANNNI axis ($K_1=K_2=K$) means it suffices to describe it's right side. 

The $AP/FE$ boundary AB ($K_1+K_2=-J$ ) is the line of frustration (disorder line) while the $AP/FI$ and $FE/FI$ boundaries BD and BC have coexisting frustrated and (anti)ferromagnetically ordered sub-lattices. Ground state entropy remains finite and fixed along each boundary, with  B being a discontinuous disorder point of maximum entropy $s = \frac{1}{2}\log 3$. $AP$ phase is four-fold degenerate with repeating sequence $\uparrow\uparrow\downarrow\downarrow$, $\uparrow\downarrow\downarrow\uparrow$, $\downarrow\downarrow\uparrow\uparrow$ or $\downarrow\uparrow\uparrow\downarrow$. The two-fold degenerate $FE$ phases in the second and fourth quadrants have sequences  $\uparrow\uparrow\downarrow\uparrow$ or $\downarrow\uparrow\uparrow\uparrow$ and  $\uparrow\uparrow\uparrow\downarrow$ or  $\uparrow\downarrow\uparrow\uparrow$.
  
The $4\times 4$ transfer matrix  $\mathbf{\mathcal{T}}$ is obtained by a consideration of Eq. \ref{Hamiltonian} as a sum over $N/2$ block Hamiltonians corresponding to adjacent parallelograms $1342$, $3564$, etc of Fig.\ref{fig1}(a). Its matrix elements are
\begin{eqnarray}
\langle u,v|{\mathbf{\mathcal{T}}}|w,x \rangle &=&  \exp\,\beta\left[\frac{J}{2} (u\,v +2\,v\,w +w\,x ) + ( K_{1}\,u\,v \right.\nonumber\\ &+&\left. K_{2}\,v\,x ) +\frac{ H}{2}(u+v+w+x)\right]\nonumber\\
\label{transfer matrix}
\end{eqnarray}
 $u$,$v=\pm 1$. The free energy (Massieu function) per site is 
\begin{equation}
\psi(y_1,y_2)=\frac{1}{2}\log \nu_+
\label{psi}
\end{equation}
where $(y_1,y_2)=(1/T,H/T)$ with $\nu_+$ the largest eigenvalue of $\mathcal{T}$.

 \textit{The correlation functions}. Following \cite{steph} the zero-field two-point correlation functions can be obtained by mapping the Hamiltonian in Eq. \ref{Hamiltonian}, via the spin transformation
$\sigma_i=s_is_{i+1}$
to an nn Hamiltonian with alternating couplings,
\begin{eqnarray}
\mathcal{H}&=& -J\sum_{i=1,2,..}\sigma_i-K_1\sum_{i=1,3..}\sigma_i\sigma_{i+1}-K_2\sum_{i=2,4,..}\sigma_i\sigma_{i+1}\nonumber\\
&=&\sum_{i=1,3,..}\mathcal{H}^{(1)}_i+\mathcal{H}^{(2)}_{i+1}.
\label{Hamiltonian nn}
\end{eqnarray}
Here the block Hamiltonians are
\begin{equation}
\mathcal{H}^{(\alpha)}_{i}=-\frac{J}{2}(\sigma_i+\sigma_{i+1})-K_{\alpha}(\sigma_i\sigma_{i+1})
\label{block Hamiltonian}
\end{equation}
with the resulting transfer matrices $\mathcal{T}^{(\alpha)}$. The partition function has the structure,
\begin{equation}
\mathcal{Z}=Tr[\langle \sigma_1|\mathcal{T}_a|\sigma_2\rangle\langle\sigma_2|\mathcal{T}_b|\sigma_3\rangle...\langle\sigma_N|\mathcal{T}_b|\sigma_1\rangle]
\label{partition function mapped}
\end{equation}

Appropriately matching configurations of the $\sigma$-chain to the original $s$-chain one has
\begin{equation}
\langle s_i s_{i+k}\rangle=\langle \sigma_i\sigma_{i+1}...\sigma_{i+k-1}\rangle= \Gamma(K_1,K_2,\beta;k)
\label{spin corr map}
\end{equation}

 There are in general three classes of $\Gamma$: two for even $k$, $\Gamma_{ev}$ and $\Gamma_{od}$ connecting even-even and odd-odd spins and a $\Gamma_{eo}$ for odd $k$ connecting even-odd spins. It will suffice to focus on the first two correlations here.
Inserting the required number of $\sigma_i$'s in Eq. \ref{partition function mapped} the odd-odd correlation function is obtained as
 \begin{equation}
\langle s_1s_{1+k}\rangle= \Gamma_{od} = \frac{Tr\,({\tilde{\mathcal{T}}}_{ab}^{\,k/2}\,\mathcal{T}_{ab}^{\,(N-k)/2})}{Tr\,(\mathcal{T}_{ab}^{\,N/2})}
 \end{equation}
 where $\mathcal{T}_{ab}=\mathcal{T}_{a}\,\mathcal{T}_{b}$ and $\tilde{\mathcal{T}}_{ab}=\mathcal{S}\,\mathcal{T}_{a}\,\mathcal{S}\,\mathcal{T}_{b}$ with the spin matrix $\mathcal{S}=\text{diagonal}\{1,-1\}$. Diagonalize the product matrices as $\Lambda=\mathcal{Q}^{\dagger}\,\mathcal{T}_{ab}\,\mathcal{Q}$ with real $\lambda_1>\lambda_2$ and as $\tilde{\Lambda}=\mathcal{P}^{\dagger}\,\tilde{\mathcal{T}}_{ab}\,\mathcal{P}$ with real or complex conjugate $\mu_{1,2}$.
 Set $\mathcal{A}=\mathcal{Q}^{\dagger}\,\mathcal{P}$ and let $N\to\infty$ to get
 \begin{eqnarray}
 \Gamma_{od}(K_1,K_2,\beta;k)&=&Tr\,(\mathcal{A}^{\dagger}\,\tilde{\Lambda}^{k/2}\,\mathcal{A}\,\,\mathcal{U})/\lambda_1^{k/2}\\
 &=& \sum_{i=1,2}|a_{i1}|^2\left(\frac{\mu_i}{\lambda_1}\right)^{k/2}
 \end{eqnarray}
 where $\mathcal{U}=\text{diagonal}\{1,0\}$.
 
  Shifting the spin position labels by one lattice constant interchanges  correlations, giving
\begin{equation}
\Gamma_{ev}(K_1,K_2,\beta;k)=\Gamma_{od}(K_2,K_1,\beta;k)
\label{exchange corr}
\end{equation}
It will be fruitful to estimate correlation lengths directly from the $\Gamma$ vs. $k$ plots.


\begin{figure*}[t!]
 \begin{subfigure}[b]{0.3\textwidth}
\centering
\includegraphics[width=2.2in,height=2in]{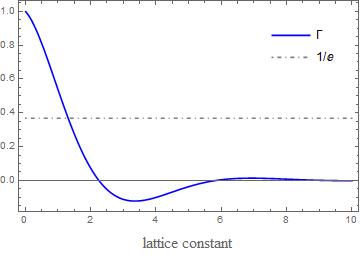}

 \caption{\small{}}
\end{subfigure}
\hfill
 \begin{subfigure}[b]{0.3\textwidth}
\centering
\includegraphics[width=2.2in,height=2in]{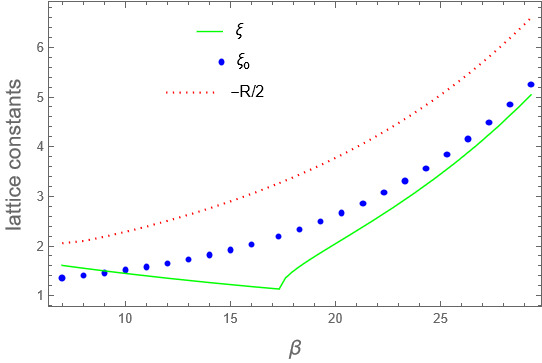}

 \caption{\small{ }}
\end{subfigure}
\hfill
 \begin{subfigure}[b]{0.3\textwidth}
\centering
\includegraphics[width=2.2in,height=2in]{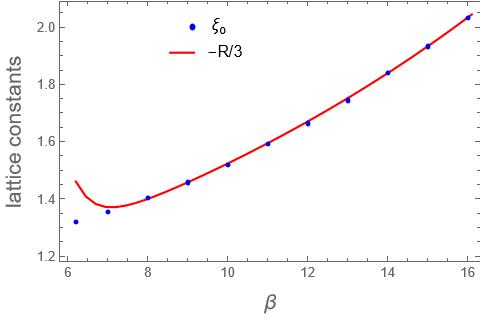}

 \caption{\small{}}
\end{subfigure}
\caption{\small{($a$) Range of order $\xi_o$(=1.323 here) in the short-range modulated order regime, with $K_1=K_2=-0.49$, $J=1$ and $\beta=12<\beta_D=17.4$. Horizontal line is at $1/e= 0.3679$. ($b$) $R$ and $\xi_o$ vary smoothly across $\beta_D$ while conventional $\xi$ has a minimum with a singular slope. ($c$) $R$ is exactly three times $\xi_o$ upto about $1.35$ lattice constants.  }}
\label{fig2}
\end{figure*}


\textit{The scalar curvature}. The invariant $R$ is obtained from the Riemannian thermodynamic metric $g_{\mu\nu}=\partial_\mu\partial_{\nu}\psi$ via standard differential geometric methods. For a $2d$ Hessian metric, as is the case here, $R$ can be expressed in terms of upto third-order derivatives of $\psi$, \cite{rupporiginal,dolan3}
\begin{equation}
R=-\frac{1}{2}\begin{vmatrix}
\psi_{,11}&\psi_{,12}&\psi_{,22}\\\psi_{,111}&\psi_{,112}&\psi_{,122}\\\psi_{,112}&\psi_{,122}&\psi_{,222}
\end{vmatrix} \times \begin{vmatrix}
\psi_{,11}&\psi_{,12}\\ \psi_{,21}&\psi_{,22}
\end{vmatrix}^{-2}.
\label{ruppdet}
\end{equation}
where the subscripts indicate derivatives with respect to $y_1=\beta$ and $y_2=\beta H$. We recall that, for example, $\psi_{12}$=$\langle\Delta m\Delta \epsilon\rangle$, $\psi_{112}$=$\langle(\Delta\epsilon)^2\Delta m\rangle$, etc where $m$ and $\epsilon$ are the magnetization and energy per lattice site. The second moments $\psi_{22}=\sigma_m^2=T\chi_T$ and $\psi_{11}=\sigma_\epsilon^2=T^2 C_H$ where $\chi_T$ and $C_H$ refer to the susceptibility and specific heat.

 Reflection symmetry of the Hamiltonian implies that in zero-field  the quantities $\psi_{12}$, $\psi_{112}$, $\psi_{222}$ vanish at all temperatures and the determinant of $g_{\mu\nu}$ reduces to $g= \psi_{11}\psi_{22}$. The zero-field curvature reduces to a simple form 
\begin{eqnarray}
R_0&=&\frac{1}{2}\partial_\beta\log\psi_{22}\left(\frac{\partial_{\beta}\log \psi_{11}-\partial_{\beta}\log \psi_{22}}{\psi_{11}}\right)\\
&=& \frac{1}{2}\frac{\alpha_m\,\rho_{\epsilon m}}{\sigma_\epsilon^2}
\label{curvature zero}
\end{eqnarray}
where we define the statistical quantities
\begin{eqnarray}
\alpha_\epsilon &=& \partial_\beta\psi_{11}/\psi_{11} = \langle(\Delta\epsilon)^3\rangle/\langle(\Delta\epsilon)^2\rangle\nonumber\\
\alpha_m &=& \partial_\beta\psi_{22}/\psi_{22}= \langle(\Delta m)^2\Delta\epsilon\rangle/\langle(\Delta m)^2\rangle\nonumber\\
\rho_{\epsilon m} &=& \alpha_\epsilon-\alpha_m.
\end{eqnarray}
The suggestive form of Eq. \ref{curvature zero} reveals an interplay of specific third order and second order fluctuation moments that principally governs the response of $R_0$ in critical as well as non-critical regimes, including its divergence, convergence to finite values, sign change, etc.

 Energy fluctuations $\psi_{11}$ typically decay everywhere to zero in the same manner as the singular free energy $\psi_s$. 
 The spin fluctuation moment $\psi_{22}$ diverges in the $FE$ and $FI$ regimes such that the fluctuation determinant $g$ remains either constant or divergent. Here the correspondence $R\sim\xi^d$ works very well both in the critical and non-critical regimes as we shall show. For example,  $R_0$ in Eq. \ref{curvature zero} fits well with with the two-scale factor universality relation, 
  \cite{rupporiginal, stauffer}. On the other hand $\psi_{22}$ decays to zero or to small values in the $AP$ phase or its boundary such that the fluctuation determinant $g$ vanishes in either scenario. Here the interpretation of $R$ would need some refinement, as we shall discuss.

\textit{Correlations across the disorder point}. Focus first on the interior of $\triangle AOB$ in Fig. \ref{fig1}(b). It is sufficient to consider only the ANNNI axis ($-0.5<K<0$) here since rest of the region behaves similarly. The nn interaction here is just strong enough to render a stable ferromagnetic ground state. 
At higher temperatures the entropy stabilises nnn coupling effects leading to the emergence of a short-ranged modulated order. 
The temperature $\beta_D$ below which the correlation decay changes from monotonic to oscillatory is called the {\textit{disorder point of the first kind}}, \cite{steph,lieb,selke}.
Notably, owing to a change of $\mu_i$'s from real to complex conjugate, it is generally agreed that here the correlation length $\xi^{-1}=\ln |\lambda_1/\mu_1|$ undergoes a minimum, with a singularity in the slope of the $\xi$ vs. $\beta $ curve, \cite{steph,selke,lieb}.

Physically it is not obvious how correlation length ought to decrease or its slope undergo a singularity as the temperature is lowered towards a ferromagnetic ground state. Arguments to the effect that here the nn and nnn interactions `cancel' each other, \cite{plini}, are difficult to follow since such a cancellation already takes place in the ground state across the $FE/AP$ boundary.  
Instead, we propose to examine the applicability of the conventional definition of $\xi$, valid for large $k$, near $\beta_D$ where it is typically of order unity.

Underlying the assumption of large $k$ is the condition that the decay of correlations assume an exponential character, $ \Gamma\sim f(k)e^{-k/\xi}$. 
For the $K_i\text{'s}>0$, this is already achieved at small lattice constants, much like for the ferromagnetic Ising model. However, this is not the case near the disorder point where the correlations becomes practically negligible for $k\geq 2$.

Here we adopt an operational estimate of the correlation length as the `range of order' $\xi_o$ at which the correlation diminishes  by an exponential factor of its  maximum value of unity, $\Gamma(k=\xi_o)=1/e$.  By design, it will match with the standard definition of $\xi$ for large $k$, whether the decay is monotonic or oscillatory. For the latter case we take into consideration the envelope of oscillations. However, the envelope may not be a useful construct for the oscillatory decay near $\beta_D$ since the initial drop in correlation is quite steep there. For such cases we obtain $\xi_o$  by intercepting the $1/e$ line directly with the oscillatory correlation function as demonstrated in Fig. \ref{fig2}(a). 

Figure \ref{fig2}(b) shows that unlike conventional $\xi$ the range of order $\xi_o$ increases smoothly across $\beta_D$, thus also bypassing the issue of slope  singularity.
Interestingly, our take on the issue has been motivated by observations of a smooth variation of $R$ across the disorder point as shown in Fig. \ref{fig2}(b). It turns out, $R$ encodes $\xi_o$ surprisingly well. 
Thus, for each point interior to $\triangle AOB$ of Fig. \ref{fig1}(b) the ratio $-R/\xi_o$ remains fixed at between $2$ and $3$ upto the disorder point $\beta_D$, beyond which it asymptotes to $2$, much like for other Ising based models, \cite{ruppmag, belucci}. Close to the line of frustration AB we find that for $\beta<\beta_D$ the curvature remains {\textit{exactly}} $3$ times $\xi_o$ down to almost a unit lattice size, see Fig. \ref{fig2}(c). 

Remarkably, the thermodynamic $R$ consistently anticipates an $operational$ estimate of correlation length in a far-from-critical regime at length scales where the conventional, non-thermodynamic measure of $\xi$ needs to be re-examined. 

\begin{figure}[!t]
\centering
\includegraphics[width=2.3in,height=2in]{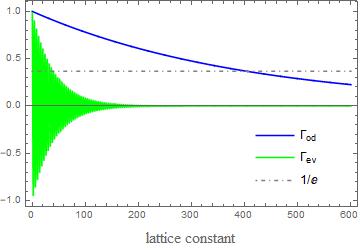}
\caption{\small{Plot showing coexistence of oscillatory and monotonic correlations $\Gamma_{ev}$ and $\Gamma_{od}$ in the $FI_A$ parameter space, with $(K_1,K_2)=(0.4,-1.2)$ and $\beta=7.5$.  Here, $\xi_{ev}<\xi_{od}$ with $R\sim\xi_{ev}$.}}
\label{fig3}
\end{figure}

\textit{Co-existence of scales in the ferrimagnetic phase}.
The $FI$ regime is qualitatively different from other phases in terms of it fluctuating substructures. For the $FE$ and $AP$ parameter regions the correlations $\Gamma_{ev}$ and $\Gamma_{od}$ become similar at low temperatures (high $\beta$) with both sub-lattices being either ferro- or antiferro-magnetic. Similarly the correlation lengths too become equal, $\xi_{ev}=\xi_{od}$, irrespective of the relative strengths of $K_1$ and $K_2$. The $FI$ regime, on the other hand, is characterised by a monotonic $\Gamma_{od}$ and an oscillatory $\Gamma_{ev}$ each with a separate correlation length, $\xi_{ev}\sim e^{-2(K_2+J)\beta}$ and $\xi_{od}\sim e^{2K_1\beta}$. 

\begin{figure}
\centering
\begin{subfigure}[b]{0.35\textwidth}
   \includegraphics[width=1\linewidth]{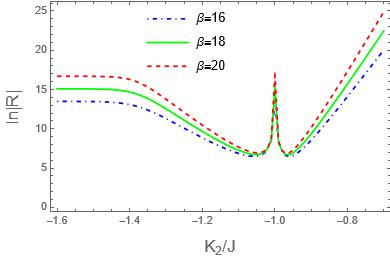}
   \caption{}
   \label{} 
\end{subfigure}
\begin{subfigure}[b]{0.35\textwidth}
   \includegraphics[width=1\linewidth]{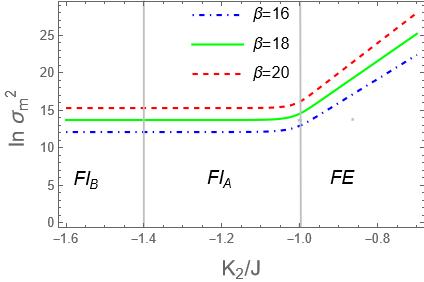}
   \caption{}
   \label{}
\end{subfigure}
\label{}
\caption{\small{Isothermal plots of the logarithm of $R$ in ($a$) and of spin fluctuation moment $\sigma_m^2=T\chi_T$ in ($b$) for $K_2$ ranging across the $FE$, $FI_A$ and $FI_B$ parameter values, with $K_1=0.4$. Overall $R$ characterizes phase changes better than the susceptibility.}}
\label{fig4}
\end{figure}


A further sub-classification of the $FI$ region is suggested by the two correlation lengths. Above the line BM at $ K_1+K_2=-J $ (see Fig. \ref{fig1}(b)) we heve $\xi_{od}>\xi_{ev}$ while $\xi_{ev} > \xi_{od}$ below BM. We shall label the former sub-region as $FI_A$ and the latter as $FI_B$. Figure \ref{fig3} demonstrates a typical correlation scenario in the $FI_A$ parameter space.

$R$ demonstrates a strong connect to the changing mesoscopics in this region. In the $FI_A$ sub-region it tracks the antiferromagnetic correlations along the even sub-lattice, with $R\sim 2\xi_{ev}$. In $FI_B$ it tracks the ferromagnetic correlations along the odd sub-lattice, with $R\sim 2 \xi_{od}$. 
Notably, in both the sub-regions the thermodynamic curvature is informed by the weaker of the two orderings.  
A plausible physical picture suggests that geometry responds to the overall order in the chain which, in turn, is tied to the sub-lattice that orders last.
The sub-lattice with a stronger ordering (larger $\xi$) will present itself as a rigid (anti)ferromagnetic background against which the slower ordering proceeds.

Figure \ref{fig4} demonstrates an overall superiority of $R$ over $\chi_T$ at characterizing the phase structure in the $FE/FI_A/FI_B$ region. For $K_1=0.4$  $R$ signals the $FE/FI$ transition at $K_2=-1$ as well as the $FI_A/FI_B$ cross-over at $K_2=-1.4$ while $\chi_T$ signals only the $FE/FI$ transition. The cross-over is also signalled by $\psi_{11}$ (not shown in figure) which scales as $\psi_s^{-1}$ .


Incidentally, the $FI_A$ region is an instance of curvature representing the anti-ferromagnetic correlations. In other known cases it asymptotes to small value, \cite{belucci}.

\begin{figure}[!t]
\centering
\includegraphics[width=2.5in,height=2.5in]{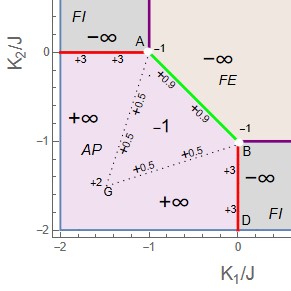}
\caption{\small{Asymptotic (ground state) values of $R$ in the antiphase ($AP$) region and surroundings. }}
\label{fig antiphase}
\end{figure}

\textit{Antiphase and its boundary}.  As mentioned earlier,  in the antiphase region and its associated boundaries the fluctuation determinant  $g$ shrinks to zero as the temperature is lowered. Here $R$ does not scale asymptotically as the inverse of $\psi_s$ since  the numerator of Eq. \ref{curvature zero} is no more asymptotically constant but decays to zero albeit never faster than $\psi_s$.
In this region we shall not make any attempt to associate $R$ with the correlation length of the physical system. Rather, we explore its usefulness in characterizing the nature of underlying statistical interactions.

In Fig. \ref{fig antiphase} the ground state $R$ changes discontinuously across $FE/AP$ and $FI/AP$ boundaries. It limits to $+0.9$ along the line of frustration AB and is a constant $-1$ on A and B, reflecting the discontinuity in the ground state entropy.
Along BC, where the even sub-lattice is antiferromagnetic, $R$ limits to +3.
Within the $AP$ phase it asymptotes to $-1$ everywhere inside $\triangle$GAB.
On the edges GB, GA given by $\omega_{1,2}\equiv K_{1,2}-3(K_{2,1}+J)=0$ 
it uniformly asymptotes to $+0.5$ and, except at point G where it is $+2$. 
 Exterior to $\triangle GAB$ curvature diverges to $positive$ infinity as $e^{\,\omega_{1,2}\beta}$.

Notably, from the expression for $R_0$ in  Eq. \ref{curvature zero}, it is only on the frustration  points that $\alpha_m$ decays like $\psi_s$ to render a constant $R$.
Everywhere else $R$ approaches a constant or diverges depending on whether the speed of decay of $\rho_{\epsilon m}=\alpha_\epsilon-\alpha_m$ matches or trails that of $\psi_s$.

Drawing on interpretations motivated by the lattice gas analogy, \cite{belucci,ruppajp}, we may think of ferromagnetic coupling as `statistically attractive' as it increases the chances of particles bunching up in adjacent cells and the antiferromagnetic coupling  as statistically repulsive which discourages clustering of particles.  In these terms, a large negative $R$ accords well with the `attractive' $FE$ and $FI$ regimes where the determinant of fluctuations $g$ remains non-zero and $R\sim\xi$.
A small $|R|$ is linked to weak repulsive interactions or solid-like phases where mutual avoidance of constituent atoms governs ordering at small scales, \cite{rupprep}. This too fits well with our observations of small $|R|$ within $\triangle GAB$ and along the partially antiferromagnetic boundary BD. Finally, a divergent positive curvature is seen in the Fermi gas where mutual $exclusion$ (strong statistical repulsion) governs ordering, \cite{ruppajp}. Possibly, $R\to+\infty$  beyond $\triangle GAB$ is similarly suggestive of the strength of antiferromagnetic nnn coupling crossing a threshold to become `strongly repulsive'. Of course such tentative associations require further analysis.

{\textit{Conclusion}}. 
 A key message in this work is that tools of thermodynamic geometry can help open up a top-down channel to extract meaningful information about underlying microscopics that remains hidden within thermodynamics. We demonstrate a significant role of thermodynamic curvature in unearthing such meso-scale insights in a model of competing interactions. It is hoped that our results will encourage workers to include $R$ as a standard thermodynamics based tool to complement their studies of physical systems. 

More work is needed to arrive at a quantifiable interpretation of $R$ where the fluctuation determinant $g$ vanishes. Third order statistical objects, including some introduced here could prove useful in this context.
Future research includes ANNNI model in non-zero field and mean field approximations of  models of frustration. 
\\\\
We thank Vikram Patil for help with computational resources.



\end{document}